\documentclass[aps,twocolumn,groupedaddress,showpacs,preprintnumbers]{revtex4}
\usepackage[dvips]{graphicx}
\usepackage[]{caption}
\usepackage{amsmath}
\usepackage{amssymb}
\begin{document}
\bibliographystyle{prsty}
\title{Perturbations in a Holographic Universe and in Other Stiff Fluid Cosmologies}
\author{T.~J. Battefeld $^{1)}$\footnote{battefeld@physics.brown.edu} and
D.~A.~Easson $^{2)}$\footnote{easson@physics.syr.edu}}
\affiliation{$^{1)}$ Physics Department, Brown University,
  Providence, RI 02912, USA.\\
$^{2)}$ Department of Physics, Syracuse University, Syracuse, NY 13244-1130, USA.}
\date{\today}
\preprint{hep-th/0408154}
\preprint{BROWN-HET-1422}
\begin{abstract}
We examine the generation and evolution of perturbations in a universe dominated by a fluid with stiff equation
of state $p=\rho$. The recently proposed Holographic Universe is an example of such a model.
We compute the spectrum of scalar and tensor perturbations, without relying on a microphysical description
of the $p=\rho$ fluid. The spectrum is scale invariant
deep inside the Hubble horizon. In contrast, infrared perturbations that enter the Hubble horizon during the stiff fluid dominated
(holographic) phase yield oscillatory and logarithmic terms in the power spectrum. We show that vector
perturbations grow during the
stiff fluid dominated epoch and may result in a turbulent and anisotropic Universe at the end of the holographic phase.
Therefore, the required period of inflation following the holographic phase cannot be much shorter
than that required in standard inflationary models.
\end{abstract}
\maketitle
\section{Introduction}
Over the last few years many attempts have been made to merge string theory with cosmology. These models incorporate certain
aspects of string theory like extra dimensions, branes, T-duality etc.,
and often drastically change our picture of the early Universe. Some popular examples are
the K\!KLM\!MT model \cite{Kachru:2003sx},
the Cyclic/Ekpyrotic model \cite{Khoury:2001wf, Steinhardt:2001vw},
Brane Gas Cosmology~\cite{Brandenberger:1988aj,Alexander:2000xv,Brandenberger:2001kj}, Randall-Sundrum phenomenology
\cite{Randall:1999ee,Randall:1999vf}
and the Pre-Big-Bang scenario \cite{Veneziano:1991ek,Gasperini:1992em}~\footnote{For reviews see
e.g.~\cite{Lidsey:1999mc,Easson:2000mj,Carroll:1999iy,Easson:2001re,Quevedo:2002xw,Gasperini:2002bn,Brax:2003fv,Battefeld:2004xw}.}.
Ultimately, these models must make contact with experimental observation; one of the most demanding being
the observation of a nearly scale invariant spectrum of density fluctuations \cite{Spergel:2003cb}.
So far, these models rely on inflation (in some form or another) in order to generate such a spectrum.

Recently, a radical new model based on the holographic principle of 't Hooft and Susskind \cite{'tHooft:1993gx,Susskind:1994vu}
was proposed by Banks and Fischler (BF) \cite{Banks:2001px,Banks:2002fe,Banks:2003ta,Banks:2004vg,Banks:2004cw}.
In this model the initial state of the Universe is described by a stiff fluid (with equation of state $p=\rho$) that saturates
the holographic covariant entropy bound. The model conforms with standard cosmology
(obeys the Einstein equations with Friedmann-Robertson-Walker metric) with this unusual form of matter.
While we focus on the Holographic Universe model, a number of authors have considered cosmological aspects of such a
stiff fluid in the literature and our conclusions generalize to those models as
well~\cite{Fernandez-Jambrina:2004gj}-\cite{Gleiser:1987jp},~\footnote{The idea of a stiff fluid present in the early Universe
can probably be attributed to Zeldovich. For a simple analysis of the perturbation spectra see e.g.,~\cite{Grishchuk} and the references
therein.}.
The microphysical interpretation of this fluid is a dense gas of black holes \cite{Banks:2004vg} which is
argued by BF to emerge naturally from a quantum-gravity regime.

It is claimed that, during the holographic phase, a scale invariant spectrum of perturbations is generated inside the
Hubble horizon. Thereafter, it is inflated to super-Hubble scales during a short
burst of inflation \cite{Banks:2004vg}. A careful calculation of the spectrum and an implementation
of the inflationary phase needs to be performed
\footnote{See however arguments in favor of a scale invariant spectrum based on symmetry in \cite{Banks:2003ta}.}.

In this article, we address the first of these issues and compute the spectrum of scalar and tensor
perturbations inside the
Hubble horizon. The spectrum originates from sub-Hubble ultraviolet (UV) perturbations, and
infrared (IR) perturbations that crossed
the Hubble radius during the holographic phase. We find that UV perturbations do indeed yield a scale invariant
spectrum; however, IR perturbations have oscillatory and logarithmic corrections to scale invariance.

Vector perturbations (VP) are found to grow during the holographic phase. We argue that the Universe
will become turbulent and anisotropic if the $p=\rho$ fluid dominates for a long enough time. We speculate that it might
be this feature that puts an end to the holographic phase, since the Universe is no longer describable by
an effective $p=\rho$ fluid.

Our main conclusion is that due to the imprints of IR perturbations, a rather long phase of inflation is required to make
the holographic model consistent with observations.
Although the holographic cosmology proposal does not eliminate
the need for a period of cosmological inflation, there is hope
that it will help resolve the initial singularity~\cite{Banks:2001px,Veneziano:2003sz}.
\section{Background}
We work with mostly negative signature $(+,-,-,-)$, scale factor $a(\eta)$, conformal time $\eta$ and
we assume a flat, Friedmann-Roberston-Walker Universe, so that
\begin{eqnarray}
ds^2&=&a^2d\,\eta ^2 -a^2\delta _{ij}d\,x^id\,x^j\,.
\end{eqnarray}
The matter content is modelled by an ideal fluid
\begin{eqnarray}
T^\alpha _{\,\,\beta}=(\rho +p)\,u^{\alpha}\,u_\beta -p\delta ^{\alpha}_{\beta}\,,
\end{eqnarray}
where $p$ is the fluid's pressure, $\rho$ is its energy density, and $u^{\alpha}$ is its
four velocity satisfying $u^{\alpha}u_\alpha=1$ (given in a comoving frame by $(u_\alpha)=(a,0,0,0)$).
The unperturbed Einstein equations read
\begin{eqnarray}
G^\alpha _{\,\,\beta}=\frac{1}{k_4^2}T^\alpha _{\,\,\beta}\,,
\end{eqnarray}
where $k_4^{-2}=8\pi G$, so that
\begin{eqnarray}
2\frac{a^{\prime\prime}}{a}-\left(\frac{a^\prime}{a}\right)^2&=&-\frac{a^2p}{k_4^2}\,,\\
3\left(\frac{a^\prime}{a}\right)^2&=&\frac{a^2\rho}{k_4^2}\,,
\end{eqnarray}
yielding the well known relation $\rho\sim a^{-3(1+w)}$ for the equation of state $p=w\rho$.
In the above and throughout, prime denotes differentiation with respect to conformal time $\eta$.

The Holographic Universe model \cite{Banks:2004cw,Banks:2004vg,Banks:2003ta,Banks:2002fe,Banks:2001px}
is characterized by an equation of state parameter $w=1$ so that
\begin{eqnarray}
a(\eta)&=&\sqrt{\eta}\,, \label{bgs}\\
\rho(\eta)&=&\rho _0 \eta ^{-3}\,.
\end{eqnarray}
The microphysical interpretation of such a fluid is involved, but is not required in the following analysis~\footnote{Our analysis
assumes that the techniques of local field theory
are applicable to describe the origin of fluctuations.
This may not be justified for a theory based on
holography. We thank T.~Banks and W.~Fischler for raising this point.}.
\section{Scalar perturbations}
\subsection{Metric}
In the longitudinal gauge the most general perturbed metric is given by
\begin{eqnarray}
ds^2&=&a^2\left[(1+2\Phi)d\,\eta ^2 -(1-2\Psi)\delta _{ij}d\,x^id\,x^j\right]\,.
\end{eqnarray}
Note that the two Bardeen potentials agree with the gauge invariant scalar metric perturbations -- there is no residual
gauge freedom. Since we do not consider anisotropic stress, the off-diagonal Einstein Equations yield
\begin{eqnarray}
\Phi=\Psi\,,
\end{eqnarray}
so that only one scalar metric degree of freedom remains.
\subsection{Matter and Metric}
Since we do not want to rely on a microphysical realization of the holographic matter, it is not straightforward
to identify the gauge invariant, scalar degree of freedom, once matter is added. Fortunately, this difficult enterprise has
already been carried out in \cite{Mukhanov:1990me}, yielding the single gauge invariant degree of freedom
\begin{eqnarray}
v=\frac{1}{\sqrt{6}l}\left(\phi _v -2z\Psi\right)\,,
\end{eqnarray}
where $l=\sqrt{8\pi G/3}$, $\phi _v$ is the velocity potential for the perturbation
of the four velocity $\delta u_i$, that is
\begin{eqnarray}
\phi _{v,i}=-\frac{2\sqrt{\beta}a^2}{c_s}\delta u_i\,,
\end{eqnarray}
and $z$ is given by
\begin{eqnarray}
z&=&\frac{a\sqrt{\beta}}{\mathcal{H}c_s}\,,\\
c_s&=&\frac{p^{\prime}}{\rho ^{\prime}}\,,\\
\beta&=&\mathcal{H}^2-\mathcal{H}^{\prime}\,,
\end{eqnarray}
where $\mathcal{H}=a'/a$.
Thus with $p=\rho$ and our background solution (\ref{bgs}) we have
\begin{eqnarray}
\beta&=&\frac{3}{4\eta ^2}\,,\\
c_s&=&1\,,\label{cs}\\
z&=&\sqrt{3\eta}\,.\label{z}
\end{eqnarray}

The advantage of $v$ lies in the fact that its action takes the simple
form of a scalar field with time dependent mass
\begin{eqnarray}
S_v=\frac{1}{2}\int \, d^{4}\!x \,\left(v^{\prime 2}-\delta ^{ij}v_{,i}v_{,j}+\frac{z^{\prime\prime}}{z}v^2\right)\,,
\label{actv}
\end{eqnarray}
and thus $v$ can be quantized in a straightforward way \cite{Mukhanov:1990me}.
\section{Scalar Perturbations in the $p=\rho$ era}
Ultimately, we are interested in the power spectrum
\begin{eqnarray}
\left| \delta _{\Phi}(\eta ,k)\right|=\left(\frac{3l^2}{8\pi ^2}\right)^{1/2}\frac{\sqrt{\beta}z}{ac_s}\frac{1}{\sqrt{k}}
\left|\left(\frac{v_k(\eta)}{z}\right)^{\prime}\right|\label{spectrum}\,,
\end{eqnarray}
where the mode functions $v_k(\eta)$ satisfy the equation of motion derived from $(\ref{actv})$ (in momentum space)
\begin{eqnarray}
v_k^{\prime\prime}+\left(c_s^2k^2-\frac{z^{\prime\prime}}{z}\right)v_k=0\,. \label{eomv}
\end{eqnarray}
For our specific background the general solution is
\begin{eqnarray}
v_{k}(\eta)=A_k\sqrt{\eta}J_0(\eta k)+B_k\sqrt{\eta}Y_0(\eta k)\,,\label{solution}
\end{eqnarray}
where $A_k$ and $B_k$ are determined by initial conditions and $J_n$ and $Y_n$ are Bessel functions. We assume the
fluctuations are seeded by quantum vacuum fluctuations and adopt the following general initial conditions that define the
vacuum at $\eta=\eta_i$
\begin{eqnarray}
v_k(\eta_i)&=&k^{-1/2}M(k\eta_i)\,, \label{initial1}\\
v_k^\prime(\eta_i)&=&k^{1/2}iN(k\eta_i)\,, \label{initial2}
\end{eqnarray}
where $N$ and $M$ obey $NM^\star +N^\star M=2$ and the conditions
\begin{eqnarray}
\left |M(k\eta_i) \right |&\rightarrow& 1\,, \label{initial3}\\
\left |N(k\eta_i) \right |&\rightarrow& 1\,, \label{initial4}
\end{eqnarray}
for $k\eta_i>\!\!\!> 1$. Note that the short wavelength behavior is generic for most choices of the vacuum.
\subsection{Short Wavelength \label{sectionscalarshort}}
For the short wavelength part of the spectrum, that is $k\eta >\!\!\!> 1$, the
Bessel functions can be expanded
\begin{eqnarray}
J_0(k\eta)&\simeq& \sqrt{\frac{2}{\pi k\eta}}\cos(k\eta-\pi /4)+\mathcal{O}\left(\frac{1}{k\eta}\right)\,,\\
Y_0(k\eta)&\simeq& \sqrt{\frac{2}{\pi k\eta}}\sin(k\eta-\pi /4)+\mathcal{O}\left(\frac{1}{k\eta}\right)\,.
\end{eqnarray}
Using (\ref{solution}) and the initial conditions (\ref{initial1})-(\ref{initial2})  the mode functions are given by
\begin{eqnarray}
v_k(\eta)=-\frac{i N}{\sqrt{k}}\cos(k(\eta -\eta_i))-\frac{M}{\sqrt{k}}\sin(k(\eta-\eta_i))\,.
\end{eqnarray}
With this, the background solution (\ref{bgs})
and (\ref{cs})-(\ref{z}) the spectrum (\ref{spectrum}) is
\begin{eqnarray}
\left | \delta _{\Phi}(\eta ,k)\right| &=&\sqrt{\frac{9}{32}}\frac{l}{\pi}\times\\
\nonumber &&\!\!\!\!\!\!\!\!\!\!\!\!\!\!\!\! \Bigg |\frac{1}{\eta ^{3/2}}\left(
M\cos(k(\eta -\eta_i))-i N\sin(k(\eta -\eta_i))\right)\\
\nonumber &&\!\!\!\!\!\!\!\!\!\!\!\!\!\!\!\!+\frac{1}{6k\eta ^{5/2} }\left(i N\cos(k(\eta -\eta_i))+M\sin(k(\eta -\eta_i))\right)
\Bigg |\,.
\end{eqnarray}
If we focus on the dominant solution $\sim \eta ^{-3/2}$, we see that the spectrum has a
scale invariant shape, since the initial conditions (\ref{initial3}) and (\ref{initial4}) give
\begin{eqnarray}
\left | \delta _{\Phi}(\eta ,k)\right| &=&\sqrt{\frac{9}{32}}\frac{l}{\pi\eta ^{3/2}}\,.\label{spectrumshort}
\end{eqnarray}
Note that this
is essentially the spectrum that is generated during inflation inside the Hubble horizon
\cite{Mukhanov:1990me}.

So far, we have only considered the short wavelength fluctuations. However, in the stiff fluid dominated phase, IR fluctuations
are entering the Hubble horizon. If a subsequent inflationary phase lasts for only a few e-foldings, these primordial
IR perturbations can leave an imprint on the CMBR.
\subsection{Long wavelength \label{sectionscalarlong}}
To discuss the long wavelength perturbations we need the general solution (\ref{solution}). To compute the
coefficients $A_k$ and $B_k$ we must choose quantum initial conditions.
However, this task is not as straightforward as in the short
wavelength case, since there are ambiguities associated to the
choice of vacuum \cite{Giovannini:2003it}. Therefore, we will consider two general choices:
\begin{enumerate}
\item The same initial conditions as for the short wavelength perturbations (\ref{initial1}) and (\ref{initial2})
\begin{eqnarray}
v_k(\eta_i)&=&k^{-1/2}\,,\\
v_k^{\prime }(\eta_i)&=&ik^{1/2}\,,
\end{eqnarray}
this is a sensible choice (e.g. for non-exotic matter
with $p\leq\rho /3$, since the oscillator (\ref{eomv}) becomes tachyonic in that case; see
\cite{Brandenberger:2002nq} for more details).
\item Realizing that $-z^{\prime\prime}/z=1/4\eta ^2>0$, we can straightforwardly quantize the oscillator
\cite{Mukhanov:1990me},
 yielding
\begin{eqnarray}
v_k(\eta_i)&=&E^{-1/2}=(2\eta_i)^{1/2}\,,\\
v_k^{\prime }(\eta_i)&=&iE^{1/2}=i(2\eta_i)^{-1/2}\,,
\end{eqnarray}
where we took the ``energy'' as $E^2=-z^{\prime\prime}/z$ in the
limit $k\eta_i<\!\!\!<1$, with $z=\sqrt{3\eta}$.
\end{enumerate}

Computing $A_k$ and $B_k$ is then straightforward. Now one can compute an exact analytic
expression for the spectrum of pertrubations (\ref{spectrum}) -- note that there is no matching condition
required, as we know the exact solution
to (\ref{eomv}). However, since the exact expression is quite cumbersome, we shall
also provide an approximate solution valid for $k\eta_i<\!\!\!<1$ and
$k\eta >\!\!\!>1$; this approximation is good for
the spectrum of perturbations (\ref{spectrum}) inside the Hubble radius
originating from IR perturbations that crossed the horizon during the holographic ($p=\rho$) phase. With $\tau\equiv k\eta$
and $\tau _i\equiv k\eta _i$ we find
\begin{widetext}
\begin{eqnarray}
\nonumber \left | \delta _{\Phi}^{(1)}(\eta ,k)\right|&=& \frac{3\sqrt{2}}{16}\frac{l}{\pi}\frac{k^{3/2}}{\tau _i^{3/2}\tau}
\left |J_1(\tau _i)Y_0(\tau _i)-Y_1(\tau _i)J_0(\tau _i)\right |^{-1}
\bigg[
\left[J_1(\tau)\right]^2\left[Y_0(\tau _i)\right]^2+\left[Y_1(\tau)\right]^2\left[J_0(\tau _i)\right]^2\\
\nonumber &&+4\tau _i\left(\left[J_1(\tau)\right]^2Y_0(\tau _i)Y_1(\tau _i)+J_1(\tau)Y_1(\tau)
\left[J_0(\tau _i)Y_1(\tau _i)+J_1(\tau _i)Y_0(\tau _i)\right]+\left[Y_1(\tau)\right]^2J_0(\tau _i)J_1(\tau _i)\right)\\
&&+4\tau _i^2
\left(
\left(J_1(\tau)Y_0(\tau _i)-Y_1(\tau)J_0(\tau _i)\right)^2+
\left(J_1(\tau)Y_1(\tau _i)-Y_1(\tau)J_1(\tau _i)\right)^2
\right)
\bigg]^{1/2}    \\
&\simeq&\frac{3}{16}\frac{l}{\pi ^{3/2}}\frac{1}{\eta ^{3/2}}
\frac{1}{\sqrt{k\eta _i}}
\left|2\cos\left(k\eta+\frac{\pi}{4}\right)\left[\gamma +\ln\left(\frac{k\eta _i}{2}\right) +2\right]-\pi
\sin\left(k\eta+\frac{\pi}{4}\right) \right|
\label{spectrumlong1}\,,\\
\nonumber \left | \delta _{\Phi}^{(2)}(\eta ,k)\right|&=&
\frac{3\sqrt{2}}{8}\frac{l}{\pi}\frac{k^{3/2}}{\tau _i\tau}
\left |J_1(\tau _i)Y_0(\tau _i)-Y_1(\tau _i)J_0(\tau _i)\right |^{-1}
\bigg[
\left[J_1(\tau)\right]^2\left[Y_0(\tau _i)\right]^2+\left[Y_1(\tau)\right]^2\left[J_0(\tau _i)\right]^2\\
\nonumber &&-2\tau _i
\left(
\left[J_1(\tau)\right]^{2}Y_0(\tau _i)Y_1(\tau _i)
-J_1(\tau)Y_1(\tau)\left[J_0(\tau _i)Y_1(\tau _i)+J_1(\tau _i)Y_0(\tau _i)\right]
+\left[Y_1(\tau)\right]^{2}J_0(\tau _i)J_1(\tau _i)
\right)\\
&&+2\tau _i^2\left(J_1(\tau)Y_1(\tau _i)-Y_1(\tau)J_1(\tau _i)\right)^2
\bigg]^{1/2}\\
\nonumber &\simeq&\frac{3}{16}\frac{l}{\pi ^{3/2}}
\frac{1}{\eta ^{3/2}}\sqrt{2}\Bigg[\left(2\cos\left(k\eta+\frac{\pi}{4}\right)\left[\gamma +
\ln\left(\frac{k\eta _i}{2}\right)+2\right]-\pi\sin\left(k\eta+\frac{\pi}{4}\right)\right)^2\\
&&\,\,\,\,\,\,\,\,\,\,\,\,\,\,\,\,\,\,\,\,\,\,\,\,\,\,\,\,\,\,\,\,\,\,\,\,\,\,\,\,\,\,
+\left(2\cos\left(k\eta+\frac{\pi}{4}\right)\left[\gamma
+\ln\left(\frac{k\eta _i}{2}\right)\right]-\pi\sin\left(k\eta+\frac{\pi}{4}\right)\right)^2\Bigg]^{1/2}
\label{spectrumlong2}\,,
\end{eqnarray}
\end{widetext}
where the superscript $^{(1,2)}$ denotes the choice of the initial conditions and $\gamma$ is Euler's constant.
Clearly, the specturm is \it not \rm scale invariant.
\subsection{Discussion}
Comparing (\ref{spectrumlong1}) and (\ref{spectrumlong2}) with (\ref{spectrumshort}) we see that IR
perturbations do not yield a scale invariant spectrum once they cross the Hubble horizon. Independent of the initial
vacuum choice, a logarithmic and an oscillatory correction are present. Furthermore, the spectrum is proportional to
$\eta ^{-3/2}$ as it was for UV perturbations in (\ref{spectrumshort}), so this is a significant effect. Note,
the factor $\sim(k\eta_i)^{-1/2}$ in (\ref{spectrumlong1})
depends on the initial conditions given on super Hubble scales -- this is not surprising since ambiguities are
well known to arise on super Hubble scales \cite{Giovannini:2003it}. The amplitude of the oscillations is
not suppressed by small numbers. This is in contrast to the case of the \it minimal trans-Planckian \rm approach to inflation
where the amplitudes are of order $H/M_{p}$ \cite{Martin:2003kp}.

Deep inside the Hubble horizon the spectrum of fluctuations is scale invariant
according to (\ref{spectrumshort}), but due to IR fluctuations entering during the
holographic phase, the spectrum becomes $k$-dependent according to (\ref{spectrumlong1}) or (\ref{spectrumlong2})
(even inside the Hubble horizon). For the oscillatory effects to be invisible in the CMBR, a period of inflation after
the $p=\rho$ phase has to last for a  sufficient number of e-foldings.

Were we would like to get an impression of the portion of the Hubble radius that must inflate
at the end of the Holographic phase
in order to make the oscillations of the IR modes unobservable in the CMBR today.
Consider an initial patch of the Universe during the holographic phase with Hubble radius $H^{-1}$.
Now let the scale factor grow by a few orders of magnitude (say $10^{3}$).
During this time the Hubble radius will grow as $H^{-1}\sim a^{3}$.
An initial quantum fluctuation with physical wavelength of order the Hubble radius
has $\lambda _{ph}(\eta)/r_H(\eta)\sim 1/a(\eta)^2$, (since $\lambda_{ph}\sim a$).
So, if inflation sets in after the scale factor has grown by the three orders of magnitude a patch with a radius of
about $10^{-6}H^{-1}$ must inflate, and not simply a patch of size $H^{-1}$.
Thus, it seems the inflationary phase in the Holographic Universe can not be much shorter than typical inflationary
phases in standard cosmological models.
\section{Vector Perturbations in the $p=\rho$ era}
Vector perturbations (VP) have recently been studied in the context of bouncing cosmologies, due to their
growing nature in a contracting Universe \cite{Battefeld:2004cd,Giovannini:2004mc}. However, if the fluid is stiff enough,
VP will also grow in an expanding Universe (as was first pointed out in \cite{Barrow1}). We will now examine
them in the context of a $\rho=p$ Universe.

The most general perturbed metric including only VP is given by \cite{Mukhanov:1990me}
\begin{eqnarray}
(\delta g_{\mu\nu})&=&-a^2
\left(\begin{array}{cc}
0&-S^i\\
-S^i&F^i_{\,\,,j}+F^j_{\,\,,i}
\end{array}\right)\,,
\end{eqnarray}
where the vectors $S$ and $F$ are divergenceless, that is $S^{i}_{\,\,,i}=0$ and $F^{i}_{\,\,,i}=0$.

A gauge invariant VP can be defined as \cite{Battefeld:2004cd}
\begin{eqnarray}
\sigma ^i&=&S^i +F^{i\,\prime} \,.
\end{eqnarray}

The most general perturbation of the energy momentum
tensor including only VP is given by \cite{bardeen1}
\begin{eqnarray}
(\delta T^{\alpha}_{\,\,\beta})&=&
\left(
\begin{array}{cc}
0&-(\rho +p)V^i\\
(\rho +p)(V^i+S^i)&p(\pi ^i_{\,\,,j}+\pi ^j_{\,\,,i})
\end{array}
\right)\,, \label{VPemt}
\end{eqnarray}
where $\pi ^i$ and $V^i$ are divergenceless. Furthermore $V^i$ is related to the
perturbation in the 4-velocity via
\begin{eqnarray}
(\delta u^\mu)=
\left(
\begin{array}{c}
0\\
\frac{V^i}{a}
\end{array}
\right)\,.\label{defV}
\end{eqnarray}
Gauge invariant quantities are given by
\begin{eqnarray}
\theta ^i&=&V^i-F^{i\,\prime}
\end{eqnarray}
and $\pi ^i$.

From now on we work in Newtonian gauge where $F^i=0$ so that $\sigma$
coincides with $S$ and $\theta$ with $V$. Note that there is no residual
gauge freedom after going to Newtonian gauge.

If we assume for simplicity $\pi ^i=0$ (no anisotropic stress),
the equations of motion for each Fourier mode $V_k$ and $S_k$ are solved by \cite{Battefeld:2004cd}
\begin{eqnarray}
S_k^i=\frac{C_k^i}{a^2}\,,
\end{eqnarray}
where $C_k^i$ is a constant and
\begin{eqnarray}
V_k^i&=&\frac{k_4^2k^2}{2(\rho +p)a^2}S_k^i\\
&\sim&\frac{k^2C_k^i}{a^{1-3w}}\label{scaling}\,.
\end{eqnarray}
Thus, in the stiff fluid dominated phase the metric perturbation is decreasing $S_k^i=\sim\eta ^{-1}$ but
$V_k^i\sim\eta$ is increasing. Nevertheless, the contribution to the perturbed energy momentum
tensor (\ref{VPemt}) is still finite and, in fact, decreasing since
\begin{eqnarray}
\nonumber (p+\rho)V_k^i&\sim& a^{-3(1+w)}a^{-1+3w}\\
&\sim&\eta ^{-2}\,.
\label{emtv}
\end{eqnarray}
\subsection{Consequences}
The interpretation of this peculiar behavior of VP is as follows: in the holographic phase there is so
much inertia in the matter sector that eddies expanding with the Universe are actually speeding up \cite{Barrow1}.
Since the energy density in the black hole fluid scales as $\rho\sim a^{-6}$ the contribution to the energy momentum
tensor due to vorticities becomes more and more significant as the Universe expands
(from Eq.~(\ref{emtv}), $(p+\rho)V_k^i \sim a^{-4}$).
This results in a turbulent Universe at the end of the holographic phase. Note that the old arguments that ruled out
turbulent cosmologies based on Nucleosynthesis, e.g.~\cite{Barrow2}, do not apply here, since the Holographic
Universe requires an inflationary phase.

Furthermore, a significant rotation always creates a significant shear
\footnote{We thank J. Barrow for useful comments on this point.}.
It seems likely that this will produce a highly anisotropic Universe after the holographic phase.
We speculate, that it is this back-reaction of VP that could put an end to the stiff fluid dominated epoch, since
after some time the simple $p=\rho$ fluid description is no longer accurate.
If this is the case, any subsequent inflationary phase has to resolve an isotropy problem, in addition to stretching UV
perturbations to super Hubble scales.

\section{Tensor Perturbations in the $p=\rho$ era}
Here we study gravitational waves in the Holographic
Universe. Consider tensor metric fluctuations \cite{Brandenberger:2003vk,Sahni:1990tx,Sahni:2001qp}
about a classical cosmological background
\begin{equation}
ds^{2}= a^{2}(\eta)\left[ d\eta^{2} - (\delta_{ij} +
h_{ij})dx^{i}dx^{j}\right]
\,.
\end{equation}
The gravitational waves are represented by the second rank tensor
$h_{ij}(\eta,\bf x\rm)$.
Expanding the Einstein-Hilbert action to second order in the
fluctuating fields yields
\begin{equation}
\delta _{2} S = \int d^{4}x \, \frac{a^{2}(\eta)}{2}
\left[(h^{i}\,_{j})'(h^{j}\,_{i})'
-\partial_{k} (h^{i}\,_{j}) \partial^{k} (h^{j}\,_{i}) \right]
\,.
\label{act2}
\end{equation}
Decomposing
$h_{ij}$ into its two polarization states  $e^{+}_{ij}$ and
$e^{x}_{ij}$ gives
\begin{equation}
h_{ij}(\eta,\bf x\rm) =h_{+}(\eta,\bf x\rm)e^{+}_{ij} + h_{x}(\eta,\bf
x\rm)e^{x}_{ij}
\,,
\end{equation}
where $h_{+}$ and $h_{x}$ are coefficient functions. Decomposing $h^{s}$, $s\in\{+,x\} $, into its Fourier modes and
defining $\mu^{s} _k =a h^{s}_k$ such that the action for both $\mu^{s} _k$ takes canonical form, we arrive at the
equations of motions
\begin{equation}
\mu ^{\prime\prime}_{k} + \left( k^{2}-\frac{a''}{a}\right)\mu_{k}=0
\,,
\end{equation}
where we suppressed the superscript $s$. Since, in the case of Holographic cosmology, $\sqrt{3} a = z$,
the above
equation has the same solution
as the equation for the scalar perturbations (\ref{eomv}). Thus the spectrum of gravitational waves behaves in
the same way as the spectrum for scalar perturbations in sections
\ref{sectionscalarshort} and \ref{sectionscalarlong}~\footnote{The presence of logarithmic terms in the spectra
of tensor modes agrees with the results found by M.~Giovannini in \cite{Giovannini:1998bp}-\cite{Giovannini:1999qj}}.
\section{Conclusion}
In this article we computed the spectrum of scalar and tensor metric perturbations in a Universe that is
dominated by a stiff fluid with equation of state $p=\rho$ (e.g., the Holographic Universe).
UV perturbations that stayed inside the Hubble horizon during
the entire holographic phase exhibit a scale invariant spectrum. However, IR perturbations
that enter the Horizon during this phase yield oscillatory and logarithmic corrections. Despite ambiguities
related to quantum mechanical initial conditions, the results seem to be robust with respect
to different choices of the initial vacuum. Because no such oscillations are observed on scales that enter the horizon today,
we conclude that the required era of inflation after the holographic phase cannot be much shorter
than it has to be in standard inflationary models of the Universe. That way, only the scale invariant
UV part of the spectrum becomes observable.

Vector perturbations exhibit growing solutions during the holographic phase. These
result in an anisotropic and turbulent Universe, if the holographic phase lasts long enough. We speculate that
it is this back-reaction of vector perturbations that will ultimately put an end to the holographic phase.

It should be noted that the holographic Universe model is still in its infancy. The implementation of the much
needed inflationary phase has
yet to be achieved. Also, a better understanding of the microphysical properties of
the black hole fluid and how it originates from an epoch of quantum gravity is desirable.
Although the holographic cosmology proposal does not eliminate
the need for a period of cosmological inflation, there is hope
that it will help resolve the initial singularity.
\begin{acknowledgments}
We would like to thank R.~Brandenberger for many useful discussions and comments on a draft of this work.
It is a pleasure to thank T.~Banks and W.~Fischler for an important personal communication and J.~Barrow for helpful discussions.
We also thank Northeasthern University for its hospitality during PASCOS 04 and NathFest where this work was
completed. DE is supported in part by the National Science Foundation under grant PHY-0094122 and by funds 
from Syracuse University.

\end{acknowledgments}

\end{document}